\documentclass[aps,prb,twocolumn,showpacs,superscriptaddress,longbibliography]{revtex4-1}

\usepackage{amsmath,amssymb,color,bm,natbib}
\usepackage{graphicx}
\usepackage{booktabs}
\usepackage{hyperref}
\usepackage{color} 
\usepackage{siunitx}
\usepackage{verbatim}

\newcommand{\je}{j_{\mathrm{e}}}
\newcommand{\tausf}{\tau}
\newcommand{\sigmaN}{\sigma_{\mathrm{N}}}
\newcommand{\rp}{\mathbf{r}_{\mathrm{p}}}
\newcommand{\rpabs}{r_{\mathrm{p}}}
\newcommand{\lambdasf}{\lambda}

\begin{document}

\title{Characterizing spin transport: detection of spin accumulation via magnetic stray field}

\author{Matthias Pernpeintner}
\affiliation{Walther-Mei{\ss}ner-Institut, Bayerische Akademie der Wissenschaften, Garching, Germany}
\affiliation{Nanosystems Initiative Munich, M\"{u}nchen, Germany}
\affiliation{Physik-Department, Technische Universit\"{a}t M\"{u}nchen, Garching, Germany}

\author{Akashdeep Kamra}
\affiliation{Department of Physics, Univeristy of Konstanz, Konstanz, Germany}
\affiliation{Walther-Mei{\ss}ner-Institut, Bayerische Akademie der Wissenschaften, Garching, Germany}

\author{Sebastian T.B. Goennenwein}
\affiliation{Institut f\"ur Festk\"operphysik, Technische Universit\"at Dresden, Dresden, Germany}
\affiliation{Center for Transport and Devices of Emergent Materials, Technische Universit\"at Dresden, Dresden,
Germany}
\affiliation{Walther-Mei{\ss}ner-Institut, Bayerische Akademie der Wissenschaften, Garching, Germany}
\affiliation{Nanosystems Initiative Munich, M\"{u}nchen, Germany}
\affiliation{Physik-Department, Technische Universit\"{a}t M\"{u}nchen, Garching, Germany}

\author{Hans Huebl}
\email[]{huebl@wmi.badw.de}
\affiliation{Walther-Mei{\ss}ner-Institut, Bayerische Akademie der Wissenschaften, Garching, Germany}
\affiliation{Nanosystems Initiative Munich, M\"{u}nchen, Germany}
\affiliation{Physik-Department, Technische Universit\"{a}t M\"{u}nchen, Garching, Germany}


\begin{abstract}
Spin transport in electric conductors is largely determined by two material parameters - spin diffusion length and spin Hall angle. In metals, these are typically determined indirectly by probing magnetoresistance in magnet/metal heterostructures, assuming knowledge of the interfacial properties. We suggest profiling the charge current induced spin Hall spin accumulation in metals, via detection of the magnetic stray field generated by the associated static magnetization, as a direct means of determining spin transport parameters. We evaluate the spatial profile of the stray field as well as the Oersted field generated by the charge current. We thus demonstrate that such a charge current induced spin accumulation is well within the detection limit of contemporary technology. Measuring the stray fields may enable direct access to spin-related properties of metals paving the way for a better and consistent understanding of spin transport therein.
\end{abstract}

\maketitle

\section{Introduction}
The field of spintronics investigates the interplay between the spin (magnetic) and charge degrees of freedom in a solid-state system~\cite{zutic_spintronics_2004, wolf_spintronics_2001}. Initial experimental techniques have focused on the electronic  or optical detection of the magnetization, where the latter is controlled or initialized via an external magnetic field. It has subsequently been realized that the magnetization direction can also be manipulated via spin-polarized charge currents utilizing the phenomenon of spin-transfer torques (STT)~\cite{Slonczewski1996,Berger1996,Ralph2008}. The physics underlying STT may be understood with reference to a simple model in which the magnetization results from the localized d-electrons while the mobile s-electrons mediate transport. Due to an exchange coupling between the s and d electrons, the mobile s-electrons experience a torque exerted by the magnetization. Reciprocally, the magnetization experiences an equal and opposite torque. This technique has successfully been employed for magnetization switching and domain wall motion, and forms the basis for a number of devices such as racetrack~\cite{Parkin2008} and STT-magnetoresistive random access memories~\cite{Akerman2005}.

\begin{figure}[t]
\centering
\includegraphics[width=80mm]{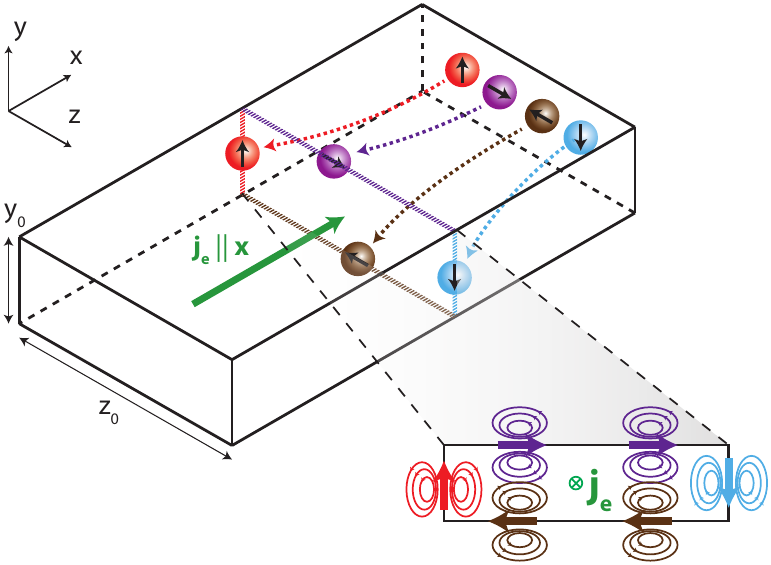}
\caption{Schematic illustration of SHE mediated spin separation and accumulation in a metallic strip. The conductor is assumed long with width $z_0$ and thickness $y_0$. A charge current density $\je$ flows along the $x$-direction. Due to spin Hall effect (SHE), the conduction electrons are scattered in different directions depending on their spin polarization: Up-spins (red; polarized along $\mathbf{\hat{y}}$), e.\,g., are deflected in the $-z$-direction, while down-spins (blue; polarized along $\mathbf{\hat{y}}$) are deflected in the $+z$-direction. This results in an accumulation of spin-polarized electrons at the surfaces of the strip. The resulting magnetization close to the edges and the magnetic fields induced by these moments are illustrated, respectively, by colored arrows and black lines. The field lines indicate the net magnetic stray field around the strip, i.\,e.~the sum of the stray and Oersted fields.}
\label{fig:Schematic}
\end{figure}

While the mechanism for spin-polarization of current relies on the conductor magnetization in the above mentioned devices, pure spin currents have also been generated and detected in non-magnetic materials, with spin-orbit interaction enabling interconversion between charge and spin currents~\cite{Dyakonov2008,hirsch_spin_1999,Saitoh2006}. Although there are a number of microscopic mechanisms contributing to this interconversion~\cite{Sinova2015}, a simple picture is provided by asymmetric scattering from impurities. An electron experiences, due to spin-orbit interaction, a spin-dependent impurity potential and scattering probability in the transverse direction (see Fig.\,\ref{fig:Schematic}). Thus, a charge flow leads to a spin current in the transverse direction and vice-versa. This phenomenon has been termed spin Hall effect (SHE) and the conversion efficiency is quantified by the so-called spin Hall angle ($\theta$). Since the spin current cannot escape the material, a spin accumulation builds up close to the conductor edges so that the diffusive backflow compensates the SHE current at the edge. This spin accumulation decays exponentially over a distance, called spin diffusion length ($\lambda$), from the interface and is well described within a diffusive transport theory~\cite{takahashi_spin_2008}.

In heterostructures comprising a magnet (F) and a non-magnetic metal (N)\footnote{Non-magnetic here shall denote metals that do not show long-range magnetic order such as ferro- or ferrimagnetism.}, the transport and magnetization electrons may be spatially separated. One mechanism for STT in these systems is via the SHE mediated accumulation of electron spins at the interface, when a charge current is driven in N. In addition to altering or moving the magnetic textures, STT also enables injection of pure spin currents into the magnetic material. This interplay between electronic and magnonic spin currents~\cite{weiler_experimental_2013} is exemplified by phenomena like spin pumping~\cite{tserkovnyak_spin_2002,czeschka_scaling_2011}, electrical spin injection~\cite{johnson_interfacial_1985}, spin Seebeck effect~\cite{uchida_observation_2008,jaworski_observation_2010,xiao_theory_2010}, and spin Hall magnetoresistance (SMR)~\cite{Nakayama2013,Althammer2013,Chen2013,Chen2016}.

Different methods for spin accumulation detection are necessary in different materials. In semiconductors, direct spatially resolved optical detection has been achieved via Kerr rotation measurements~\cite{kato_observation_2004,stern_current-induced_2006} and recently, Stamm et al. reported (non-spatially resolved) detection of the spin accumulation in metal thin films \cite{Stamm2017}.  The latter turns out to be challenging in metals due to their small electromagnetic field penetration depths and the resultig Kerr angles of the order of \SI{10e-9}{rad}. Typical techniques employed in metals rely therefore on examining an effect of the spin accumulation and constitute an indirect measurement. For example, the N thickness dependence of SMR in an F$|$N heterostructure allows inferring the spin Hall angle, but the approach relies on accurate knowledge about the interface and the interplay between the material systems~\cite{Meyer2014,Vlietstra2013}. These interfacial properties are not easily determined and vary in a wide range~\cite{weiler_experimental_2013}.

Here we suggest to detect SHE mediated spin accumulation, and thus characterize spin transport parameters, in a metallic strip by measuring the magnetic `stray' field resulting from the non-equilibrium magnetization associated with the spin-polarized electrons. While the net magnetic moment in the system vanishes, a finite magnetization is generated near the boundaries of the metal. We evaluate the ensuing stray field analytically within a simplified model as well as numerically, and find that the field is well within the detection range of the state-of-the-art sensing techniques such as NV centers\cite{Maze:2008cs,Grinolds:2013gi,Maletinsky:2012ge}, magnetic force microscopy\cite{rugar_single_2004,Taylor:2008bp}, scanning SQUID magnetometers\cite{Vasyukov:2013ed,Kirtley:2016kv}, or muon spin resonance~\cite{Leutkens2003}. We further show that the magnetic stray field of spin accumulation may exceed and can be disentangled from the Oersted field arising due to the current flow, that generates the spin accumulation via SHE, using their distinct spatial profiles. The proposed method thus enables a direct access to important spin transport properties - spin diffusion length $\lambda$ and the spin-Hall angle $\theta$ in metals - while circumventing the difficulties associated with F$|$N interfaces.

The paper is organized as follows: In Section \ref{sec:SpinAcc}, we derive the spin accumulation profile in the metallic strip (Fig.\,\ref{fig:Schematic}) and obtain an analytic expression for the magnetic stray field at large (compared to $\lambda$) distances from the surface. Section\,\ref{sec:values} discusses the spatial magnetic field distribution evaluated using the approximate analytic expression as well as numerically. In Section \ref{sec:oersted} we evaluate the Oersted field distribution generated by the charge current in the strip. We discuss the field distribution for a multilayer system in Sec.\,\ref{sec:trilayer} and demonstrate that the Oersted field can be reduced significantly by allowing a counterflow of current in an adjacent layer. We conclude with discussion of experimental issues and a short summary of our results in Section \ref{sec:summary}.

\section{Spin accumulation and magnetic stray field\label{sec:SpinAcc}}

We consider a long metallic strip with width $z_0$ and thickness $y_0$ which supports a charge current density of $\je$ driven by an electric field $E_0 \mathbf{\hat{x}}$ along its length (Fig. \ref{fig:Schematic}). The general current response in a non-magnetic conductor including SHE reads~\cite{takahashi_spin_2008}:
\begin{widetext}
\begin{equation}
 \left( \begin{array}{c}
  \mathbf{j}_{e} \\ \mathbf{j}_{sx} \\ \mathbf{j}_{sy} \\ \mathbf{j}_{sz}
 \end{array} \right)
 = \sigma_N
 \left( \begin{array}{cccc}
 1 & \theta \mathbf{\hat{x}} \times & \theta \mathbf{\hat{y}} \times & \theta \mathbf{\hat{z}} \times \\
 \theta \mathbf{\hat{x}} \times & 1 & 0 & 0 \\
 \theta \mathbf{\hat{y}} \times & 0 & 1 & 0 \\
 \theta \mathbf{\hat{z}} \times & 0 & 0 & 1
        \end{array} \right)
        \left( \begin{array}{c}
                \mathbf{E} \\  - \pmb{\nabla} \mu_{sx} / {2e} \\ - \pmb{\nabla} \mu_{sy} / {2e} \\ - \pmb{\nabla} \mu_{sz} / {2e}
               \end{array} \right),
\end{equation}
\end{widetext}
where $\sigma_N$ is the conductivity, $e (>0)$ is the electronic charge, $\theta$ is the spin Hall angle, $\mathbf{j}_{e}$ is the charge current density, $\mathbf{E}$ is the applied electric field, and $\mathbf{j}_{si}$ is the spin current density polarized in the $i$-direction ($i=x,y,z$). $\mu_{si}$ is the corresponding spin chemical potential, which obeys the diffusion equation: \cite{chen_theory_2013, Aqeel2017, fabian_semiconductor_2007}
\begin{equation}\label{eq:diffusionequation}
 \nabla^2 {\mu}_{si} = \frac{{\mu}_{si}}{\lambda^2},
\end{equation}
with the spin diffusion length $\lambda$. The boundary conditions for (\ref{eq:diffusionequation}) are vanishing spin current flow normal to all interfaces, which in the chosen coordinate system read:
\begin{equation}
 j_{si}^{y} (y = \pm y_0/2) = 0 \quad\text{and}\quad j_{si}^{z} (z = \pm z_0/2) = 0. \label{eq:boundaryconditions}
\end{equation}
Here the superscript denotes the spatial direction of spin current flow while the subscript represents the spin polarization direction. The diffusion equation (\ref{eq:diffusionequation}) with the boundary conditions [Eq. (\ref{eq:boundaryconditions})] admits the solution~\cite{Mosendz2010}:
\begin{align}\label{eq:musy}
\mu_{sy}(\mathbf{r}) & = - 2 e \theta \lambda E_0 \frac{\sinh{(z/\lambda)}}{\cosh{(z_0/(2\lambda))}},  \\
\mu_{sz}(\mathbf{r})  & =  2 e \theta \lambda E_0 \frac{\sinh{(y/\lambda)}}{\cosh{(y_0/(2\lambda))}}, \label{eq:musz}
\end{align}
where $\mathbf{r}$ is the position vector. As detailed in Appendix \ref{app:spinaccdens}, the spin accumulation density is related to the spin chemical potential by \cite{fabian_semiconductor_2007}
\begin{equation}\label{eq:spinaccdensity}
 n_{si} = \frac{\sigma_N}{2 e^2 D} \mu_{si},
\end{equation}
where $D=\lambdasf^2/\tausf$ denotes the electron diffusion constant \cite{johnson_charge_2003,takahashi_spin_2008} and $\tausf$ is the spin-flip time. The $n_{si}$ are defined as $n_{si}=n_{\uparrow}-n_{\downarrow}$, where the subscript arrows $\uparrow, \downarrow$ denote the up- and down-polarized spins for the respective quantization axes.

The spin accumulation is thus spatially localized to a region within $\sim \lambda$ from the surfaces. While the exact evaluation of the magnetic field arising from this charge-current induced magnetization requires numerics, analytical expressions can be obtained in the limit of $\lambda \ll r_p$, where $\mathbf{r}_p$ is the position vector of the point at which the magnetic field is measured. We refer to this as the `far-field limit'. Relegating the details to Appendix \ref{app:fluxdensity}, the magnetic field distribution $\mathbf{B}(\mathbf{r}_p)$ in this limit is evaluated as:
\begin{widetext}
\begin{align}\label{eq:Btot}
\mathbf{B}(\mathbf{r}_p) & = \frac{\mu_0\gamma\hbar}{8\pi} \frac{j_e \theta\tausf}{e} \left(\frac{1}{\cosh\left(\frac{y_0}{2\lambda}\right)} - \frac{1}{\cosh\left(\frac{z_0}{2\lambda}\right)}\right)\mathbf{F}(y_0,z_0;\mathbf{r}_p), \\
\mathbf{F}(y_0,z_0;\mathbf{r}_p) & =  \sum_{\sigma_{1},\sigma_{2} = \pm 1}  \frac{2(y_p-\sigma_1 y_0/2)}{(y_p-\sigma_1 y_0/2)^2+(z_p- \sigma_2 z_0/2)^2} \mathbf{\hat{y}} + \frac{2(z_p- \sigma_2 z_0/2)}{(y_p- \sigma_1 y_0/2)^2+(z_p- \sigma_2 z_0/2)^2} \mathbf{\hat{z}}, \label{eq:F}
\end{align}
\end{widetext}
where $\mu_0$ is the permeability of free space and $\gamma$ is the gyromagnetic ratio in the metal. From the expression above, we note that a high aspect ratio leads to larger stray field. Thus it is desirable to have the metal in the shape of a film.

\section{Magnetic stray field: spatial profile \label{sec:values}}

We next compute the spatial distribution of the stray field originating from spin accumulation for the example of a platinum (Pt) conductor. The material parameters employed~\cite{Meyer2014} are spin Hall angle $\theta=0.08$, electric conductivity $\sigmaN=\SI{9.52e6}{A/Vm}$, spin diffusion length $\lambda=\SI{4}{nm}$, spin flip time  $\tausf=\SI{60}{ps}$ \footnote{This value is calculated from Ref.\,\cite{Sinova2015}, Table I and Ref.\,11 therein.} and $\gamma\hbar=\mu_{\text{B}}=\SI{9.27e-24}{J/T}$. For the geometric dimensions of the Pt strip we choose $y_0=\SI{2}{nm}$ and $z_0=\SI{30}{nm}$ and we assume a current density $\je=\SI{1e10}{A/m^2}$.

\begin{figure}[tb]
\centering
\includegraphics[width=80mm]{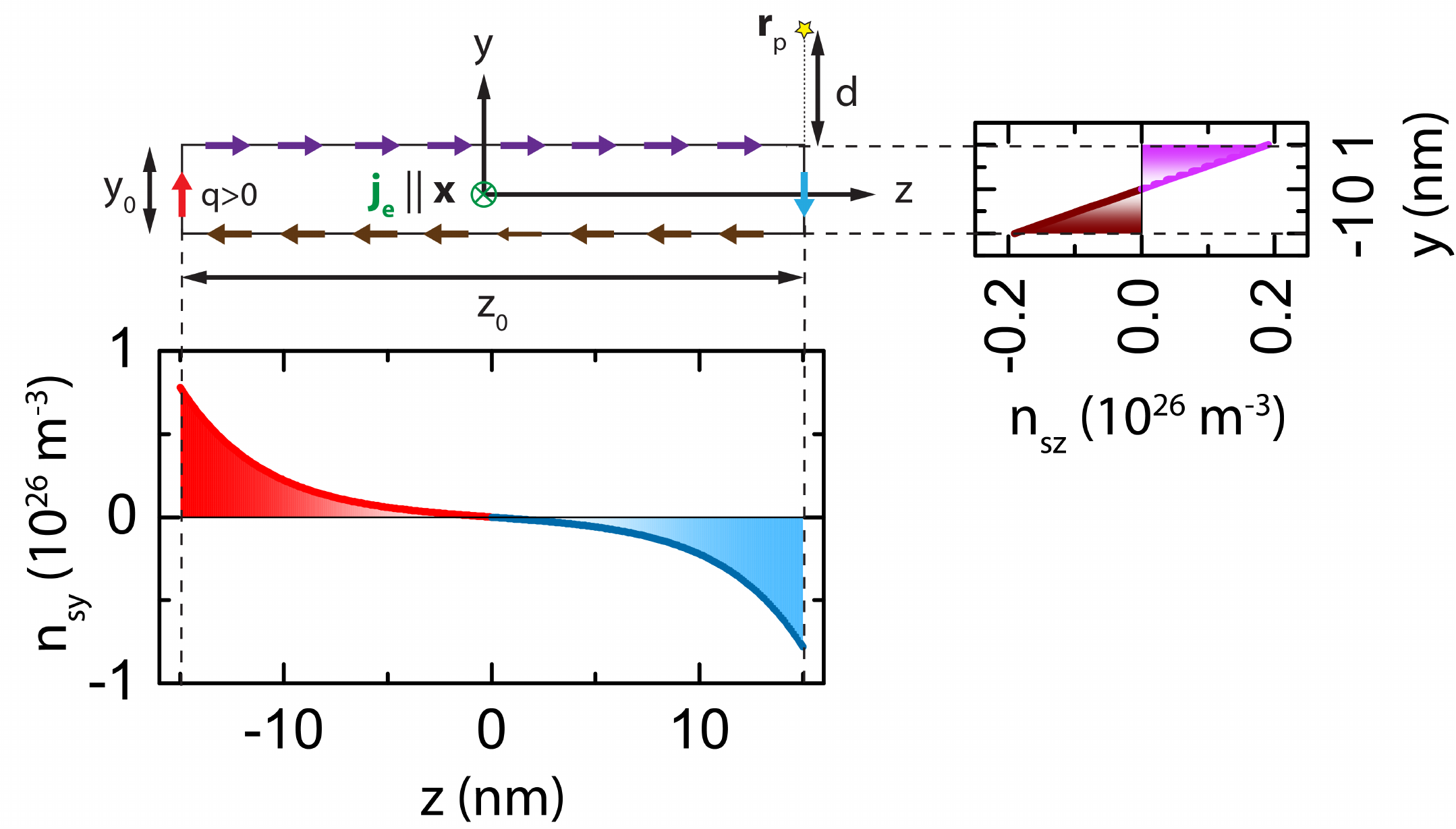}
\caption{Schematic illustration of the spin accumulation in the platinum strip, shown together with the calculated spin accumulation as a function of $y$ resp.~$z$.}
\label{fig:SingleLayerSchematicWithSpinAcc}
\end{figure}

With these material parameters, we calculate the spin accumulation at the surfaces $|n_{sy}(z=\pm z_0/2)|=\SI{7.8e25}{m^{-3}}$ and $|n_{sz}(y=\pm y_0/2)|=\SI{1.9e25}{m^{-3}}$. This corresponds to a net spin polarisation of about 0.1 percent present at the interface \footnote{Here, we have compared the calculated $n_{si}$ to the experimentally determined free electron density in platinum thin films, $n=\SI{1.6e28}{m^{-3}}$ (see Ref.~\citenum{fischer_mean_1980}).}. As evident from Eqs.~(\ref{eq:musy}) and (\ref{eq:musz}), the spin polarisation decays exponentially with decay length $\lambda$ into the body of the metal, as shown in Fig.~\ref{fig:SingleLayerSchematicWithSpinAcc}.

\begin{figure}[tb]
\centering
\includegraphics[width=80mm]{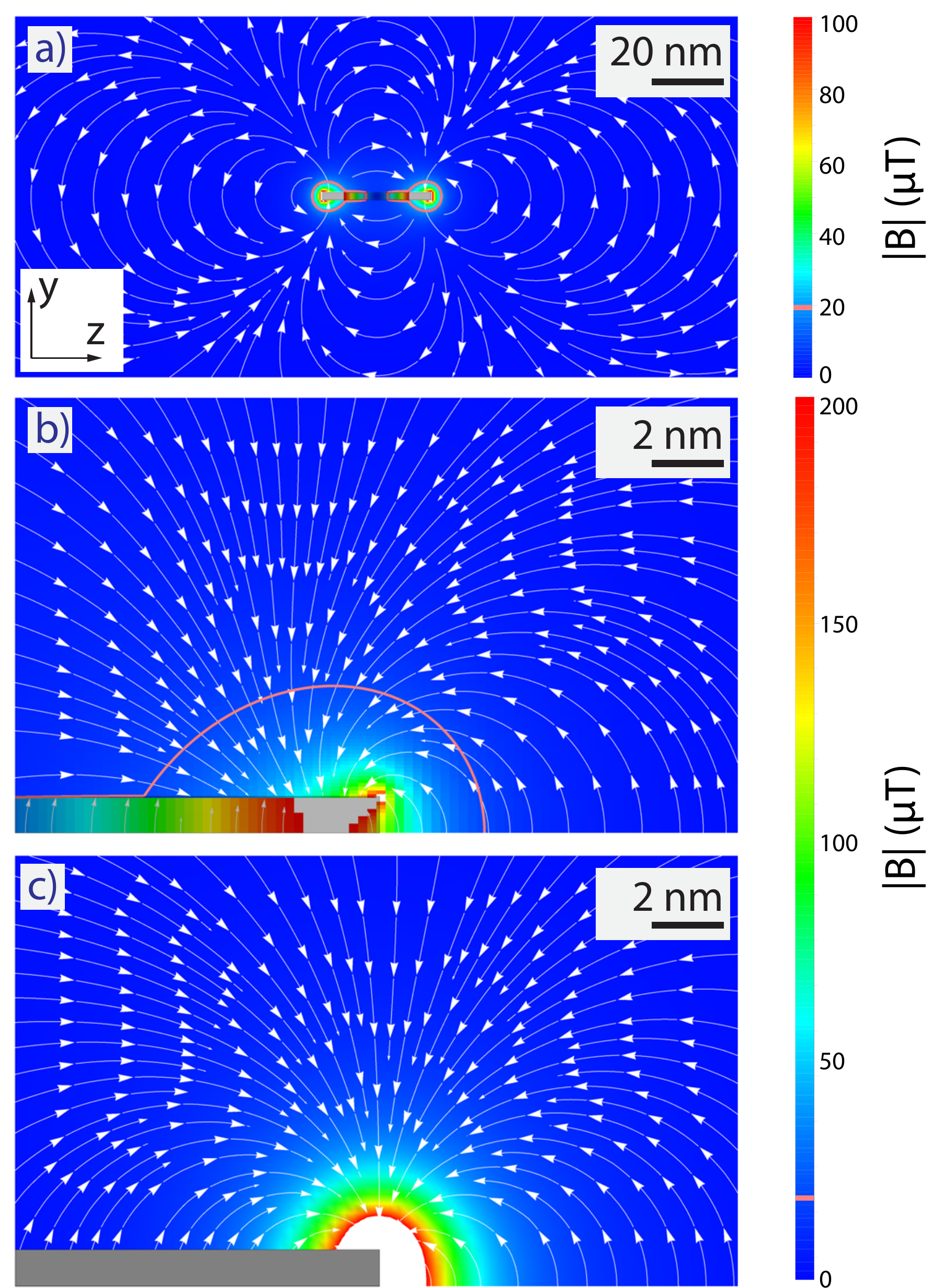}
\caption{Magnetic stray field profile $\mathbf{B}(\rp)$ of spin accumulation in the conducting strip evaluated (a) and (b) numerically as well as (c) analytically using Eq. \eqref{eq:Btot}. The white arrows indicate the magnetic field direction, the color encodes its magnitude, where white regions indicate fields above $\SI{200}{\micro T}$. The transparent (solid) gray rectangle depicts the cross-section of the metal strip for the numerical (analytical) evaluation. The pink solid line represents the $\SI{20}{\micro T}$ contour line. Panels (b) and (c) show a zoom-in around the top-right edge of the strip to compare the numerical and analytical model.}
\label{fig:StrayfieldColorPlot}
\end{figure}

The  corresponding spatial distribution of the magnetic stray field calculated numerically (see Appendix) is plotted in Fig.~\ref{fig:StrayfieldColorPlot}a. Here, the gray transparent box indicates the conductor cross-section. The stray field diverges at the edges of the strip, exceeding $\SI{20}{\micro T}$ within a radius of about $d=\SI{5}{nm}$ (Fig.~\ref{fig:StrayfieldColorPlot}a). The stray field calculated using Eq. (\ref{eq:Btot}) matches the numerical solution very well at large distances (Fig.~\ref{fig:StrayfieldColorPlot}c). Near the conducting strip, however, the approximation \eqref{eq:Btot} leads to significant errors. In the far-field limit, the stray field decays $\sim 1/\rpabs^3$.

\section{Oersted field: spatial profile \label{sec:oersted}}
\begin{figure}[tb]
\centering
\includegraphics[width=80mm]{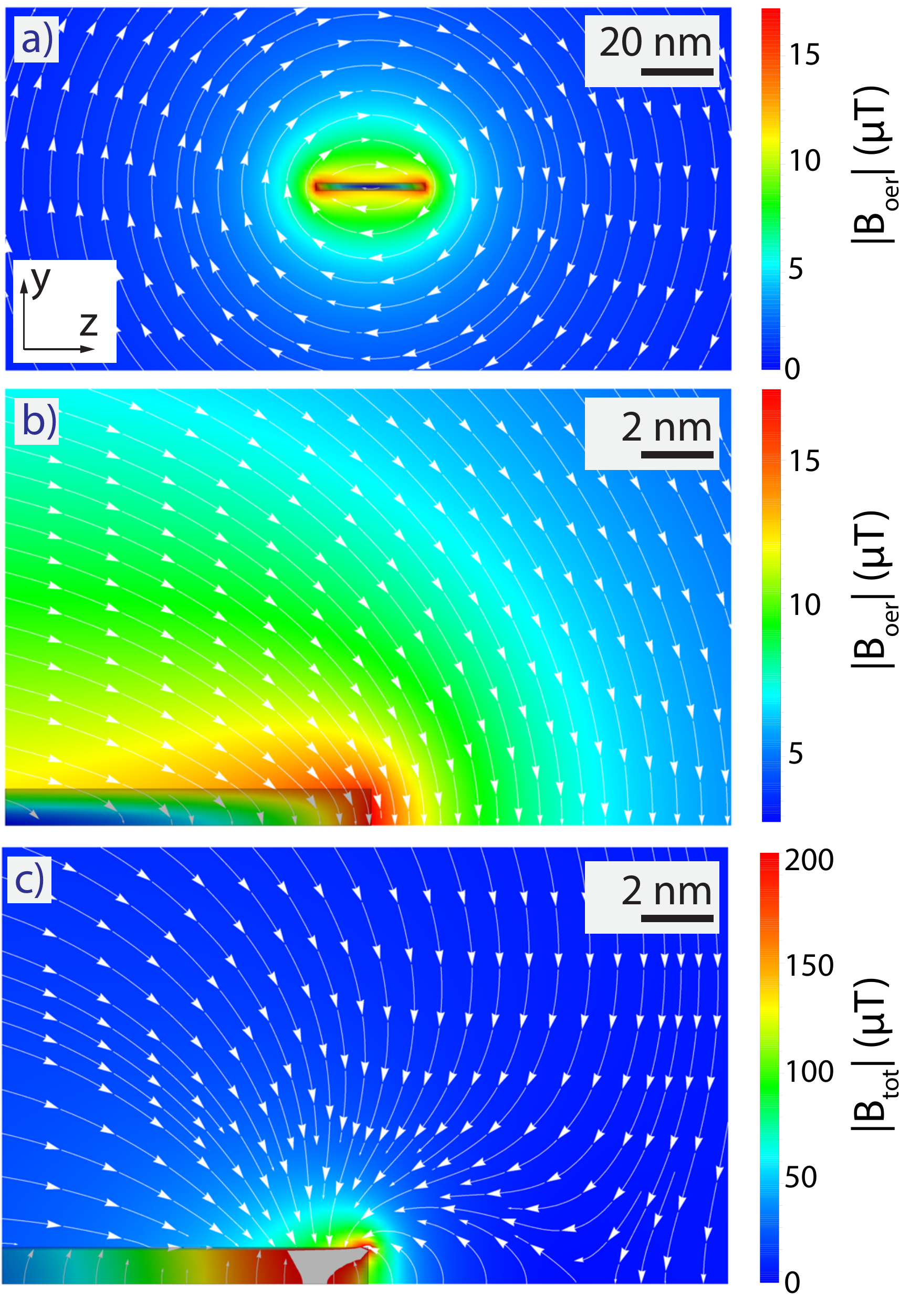}
\caption{Oersted field $\mathbf{B}_{\mathrm{oer}} (\rp)$ as a function of the sensor position $\rp$. Panel (b) depicts a zoom-in of the upper right edge and panel (c) shows the total magnetic field $|\mathbf{B}_{\mathrm{tot}}|=|\mathbf{B}+\mathbf{B_{\text{oer}}}|$.}
\label{fig:OerstedFieldSLSpatial}
\end{figure}

\begin{figure}[tb]
\centering
\includegraphics[width=80mm]{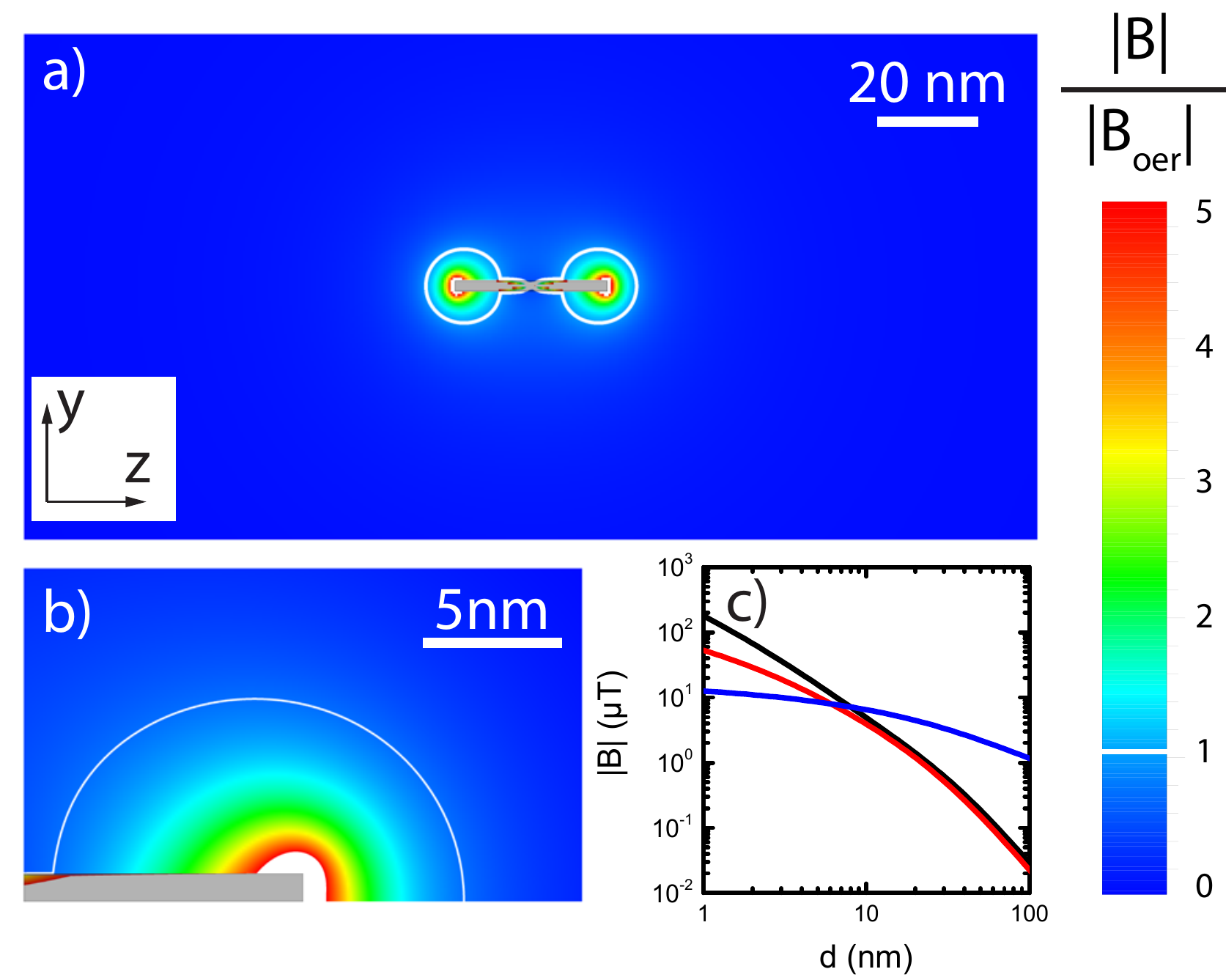}
\caption{\textbf{a.} $|\mathbf{B}|/|\mathbf{B_{\text{oer}}}|$ as a function of $y_p$ and $z_p$. The white solid line represents the $|\mathbf{B}|/|\mathbf{B_{\text{oer}}}|=1$ contour line indicating that the spin accumulation induced stray field exceeds the oersted field significantly. Areas, where $|\mathbf{B}|/|\mathbf{B_{\text{oer}}}|$ exceeds 5 are displayed in white. The gray (semi-transparent) rectangle depicts the cross-section of the metal strip. \textbf{b.} Close-up of the edge region of the strip. \textbf{c.}  $|\mathbf{B}|$  and $|\mathbf{B_\mathrm{oer}}|$ as a function of $d$ for the sensor position depicted in Fig.~\ref{fig:SingleLayerSchematicWithSpinAcc}. The solid red (black) line corresponds to the full numerical (analytical, i.e. (\ref{eq:Btot})) computation of $|\mathbf{B}|$, while the blue line depicts $|\mathbf{B_{\text{oer}}}|$. We find $|\mathbf{B}|/|\mathbf{B_{\text{oer}}}|>1$ for $d\lesssim \SI{6}{nm}$.%
}
\label{fig:SinglelayerStrayfieldColorPlotAndFarfieldLimits}
\end{figure}

Relegating the evaluation details to Appendix \ref{app:oerstedfield}, we discuss the magnetic field distribution of the Oersted field $\mathbf{B}_{\mathrm{oer}}(\rp)$ created by the charge current flow in the conductor. Figure~\ref{fig:OerstedFieldSLSpatial} shows the spatial distribution of the Oersted field around the conductor. It has its maximum of about $\SI{16}{\micro T}$ at the left and right edge of the strip. In the far-field limit, the Oersted field decays proportional to $1/\rpabs$ as expected for the far-field. Thus, at large distances the Oersted field dominates the stray field. This is also illustrated in Fig.~\ref{fig:SinglelayerStrayfieldColorPlotAndFarfieldLimits}, where the ratio $|\mathbf{B}|/|\mathbf{B_{\text{oer}}}|$ is plotted as a function of the sensor position $\rp$ including white solid line indicating $|\mathbf{B}|/|\mathbf{B_{\text{oer}}}|=1$. Nevertheless, the spatial dependence of the Oersted field significantly differs from that of the magnetic stray field of spin accumulation. Thus, using a spatially resolved magnetic field sensing technique should allow to disentangle the SHE induced stray field from the Oersted field.

\section{Trilayer geometry \label{sec:trilayer}}
In order to suppress the contribution of the Oersted field to the total magnetic field, we suggest a trilayer sample geometry where the strip consists of two conducting layers with a thin insulating layer (thickness $d_{\text{ins}}$) in between. We consider the upper layer (thickness $y_0$) to have a large spin Hall angle $\theta$, while the spin Hall angle of the lower conducting layer (thickness $y'_0$) vanishes. In the following we discuss the situation, where current flows through both conducting layers with equal magnitude but opposite signs. In the near field, the trilayer geometry reduces the Oersted field contribution. As we assume the spin Hall angle in the bottom conducting layer to be zero, the stray field of the top layer is not affected by the bottom layer. As a consequence, the ratio $B/B_{\text{oer}}$ can be increased significantly.

\begin{figure}[tb]
\centering
\includegraphics[width=80mm]{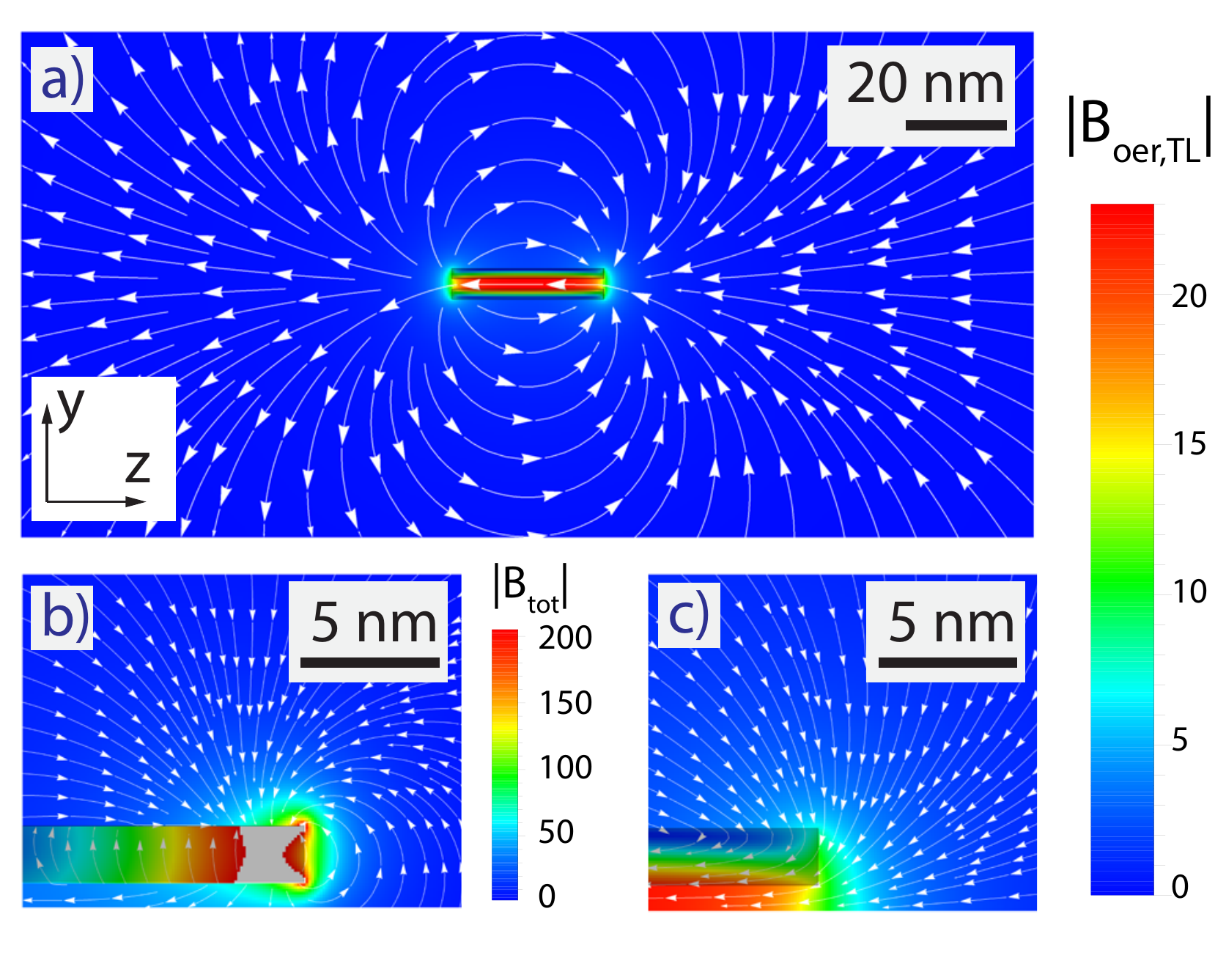}
\caption{\textbf{a.} Oersted field $\mathbf{B_{\text{oer,TL}}}$ as a function of $y_p$ and $z_p$ for the proposed trilayer sample. The semi-tranparent rectangles depict the cross-sections of the two metal strips. \textbf{b.} Total magnetic field $\mathbf{B_{\text{tot}}}= \mathbf{B}+\mathbf{B_{\text{oer,TL}}}$ close to the edge of the upper conductive strip. \textbf{c.} Oersted field of the same region for comparison.}
\label{fig:TrilayerOerstedField}
\end{figure}

\begin{figure}[tb]
\centering
\includegraphics[width=80mm]{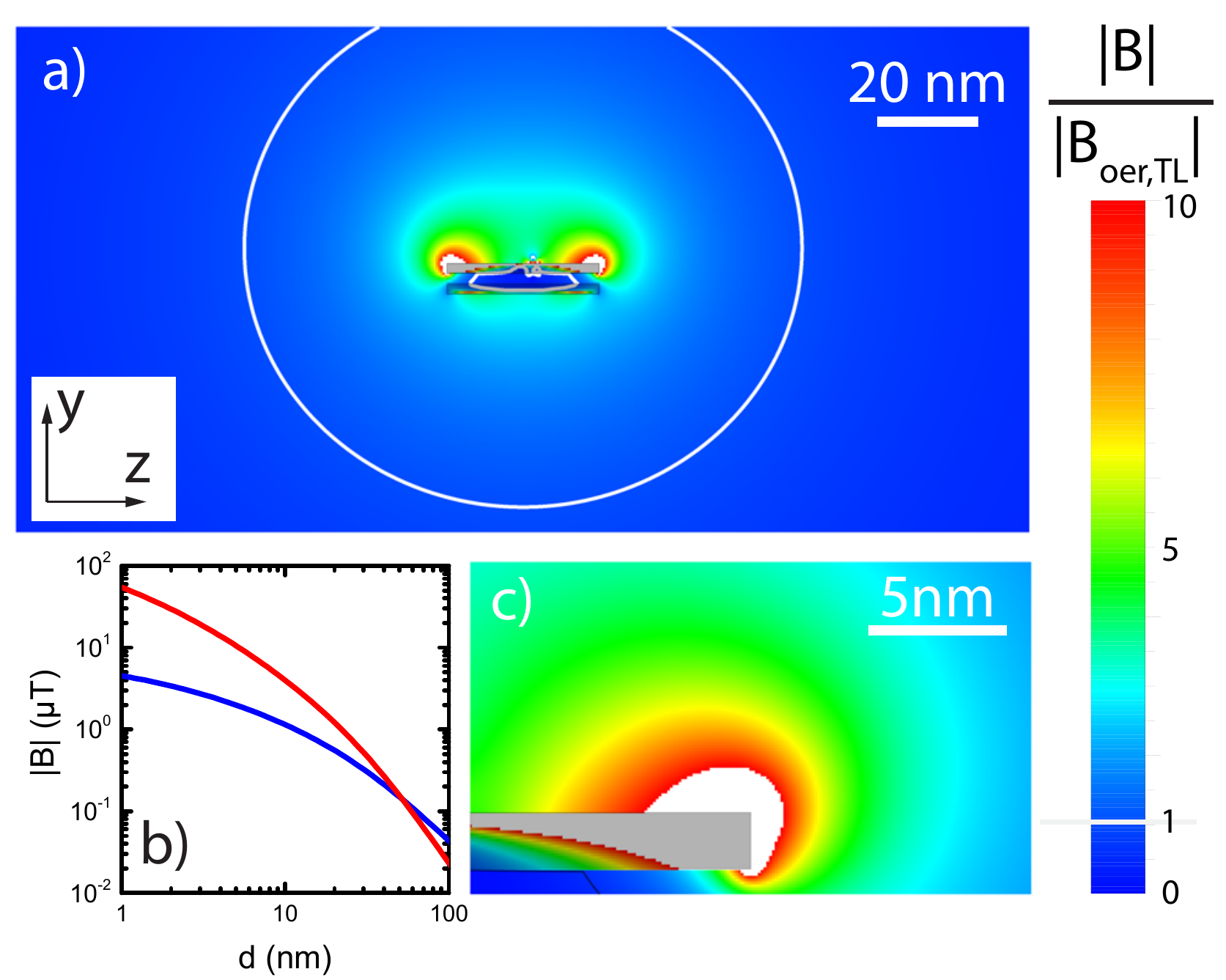}
\caption{\textbf{a.} and \textbf{c.} Field ratio $|\mathbf{B}|/|\mathbf{B_{\text{oer,TL}}}|$ as a function of $y_p$ and $z_p$ for the proposed trilayer sample. The white solid line represents the $|\mathbf{B}|/|\mathbf{B_{\text{oer}}}|=1$ contour line. Areas, where $|\mathbf{B}|/|\mathbf{B_{\text{oer}}}|>10$ are also shaded in white. The gray (semi-transparent) rectangle depicts the cross-section of the metal strip. \textbf{b.} $|\mathbf{B}|$, $|\mathbf{B_{\text{oer,TL}}}|$ and $|\mathbf{B_{\text{oer}}}|$ as a function of $d$ for the sensor position depicted in Fig.~\ref{fig:SingleLayerSchematicWithSpinAcc}. 
\textbf{c.}  $|\mathbf{B}|$  and $|\mathbf{B_\mathrm{oer}}|$ as a function of $d$ for the sensor position depicted in Fig.~\ref{fig:SingleLayerSchematicWithSpinAcc}. The solid red line corresponds to the full numerical computation of $|\mathbf{B}|$, while the blue line depicts $|\mathbf{B_{\text{oer}}}|$ of the trilayer configuration. We find $|\mathbf{B}|/|\mathbf{B_{\text{oer}}}|>1$ for $d\lesssim \SI{50}{nm}$.
}
\label{fig:TrilayerStrayfieldColorPlotAndFarfieldLimits}
\end{figure}

For a quantitative analysis, we calculate both the stray field and the Oersted field around the trilayer geometry as a function of the sensor position $\rp$. We here set $y_0=y_0'=\SI{2}{nm}$, $d_{\text{ins}}=\SI{2}{nm}$ and $z_0=\SI{30}{nm}$ \footnote{$z_0$ can be chosen large compared to $y_0$ without significantly decreasing the stray field!} and leave the material parameters unchanged. Figure~\ref{fig:TrilayerOerstedField} shows the calculated Oersted field for this trilayer geometry. Compared to the above discussed single-layer geometry (see Fig.~\ref{fig:OerstedFieldSLSpatial}), we observe a significant suppression of the Oersted field. The ratio $B/B_{\text{oer,TL}}$, plotted in Fig.~\ref{fig:TrilayerStrayfieldColorPlotAndFarfieldLimits}a, shows maxima around the edges of the top strip where the stray field clearly dominates the Oersted field. In particular, we find that the ratio of stray field and Oersted field, $B/B_{\text{oer}}$, is $5.5$ at $d=\SI{5}{nm}$ and $3.4$ at $d=\SI{10}{nm}$. Thus the contribution of the spin accumulation to the total magnetic field around the conductor is easily detectable and quantifiable in the presented geometry.

Table\,\ref{tab:values} lists the $y$-components of magnetic field $B_y$ and magnetic field gradient $\partial B_y/\partial y$ for a sample-sensor distance of $d=\SI{5}{nm}$ and $d=\SI{20}{nm}$ (cf. Fig \ref{fig:SingleLayerSchematicWithSpinAcc}).
The $d$-dependence of stray and Oersted fields is depicted in Fig.~\ref{fig:TrilayerStrayfieldColorPlotAndFarfieldLimits}b. We find that the Oersted farfield around the proposed trilayer sample decays proportional to $1/\rpabs^2$, compared to $1/\rpabs$ for the Oersted field of a single conducting layer. Besides, fig.~\ref{fig:TrilayerStrayfieldColorPlotAndFarfieldLimits}b shows the $1/\rpabs^3$-dependence of the stray field. As a consequence, in trilayer geometry, up to $d=\SI{100}{nm}$ away from the edges, the magnetic stray field exceeds the Oersted field of the two conducting layers.

\begin{table}[b]
  \begin{center}
  \caption{$y$-components of magnetic field $B_y$ and magnetic field gradient $\partial B_y/\partial y$ for a sample-sensor distance of $d=\SI{10}{nm}$ and $d=\SI{5}{nm}$ (trilayer sample geometry).}
    \begin{tabular}{cccc}
    \hline
    {}              & {}                              & \textbf{Stray field}   & \textbf{Oersted field} \\
    \hline\hline
    $d=10 {nm}$        & $B_y$                           &$\SI{-3.6}{\mu T}$       &$\SI{-1.0}{\mu T}$ \\
    {}              & $\partial B_y/\partial y$       & $\SI{460}{T / m}$           & $\SI{100}{T / m}$\\
    \hline
    $d=5 {nm}$         & $B_y$                           & $\SI{-8.8}{\mu T / m}$        & $\SI{-1.9}{\mu T / m}$\\
    {}              & $\partial B_y/\partial y$       & $\SI{1560}{T / m}$           & $\SI{282}{T / m}$\\
    \hline
    \end{tabular}%
    \end{center}
  \label{tab:values}%
\end{table}%

\section{Discussion and Summary \label{sec:summary}}

We consider magnetic force microscopy (MFM) as a potential candidate for the measurement of the stray field profile~\cite{meyer_scanning_2004,schwarz_magnetic_2008} and estimate the sensitivity required. The force acting on a MFM tip is $\mathbf{F}=(\mathbf{m}\cdot\nabla)\mathbf{B}$, where $\mathbf{m}=(0,m,0)$, $m\approx\SI{1e-13}{emu}=\SI{1e-16}{Am^2}$ is typical magnetic moment of a MFM tip~\cite{ferri_atomic_2012}. Using $\partial B_y/\partial y$ from Tab.~\ref{tab:values}, we expect a force in $y$-direction, $|F_y|$, of $\SI{46}{fN}$ (for $d=\SI{10}{nm}$) or $\SI{156}{fN}$ ($d=\SI{5}{nm}$), respectively. The state-of-the-art sensitivity concerning force measurements using MFM is about $\SI{10}{fN}$ at room temperature~\cite{jiles_magnetism_2003}. Besides, Mamin et al.~\cite{mamin_sub-attonewton_2001} have reported the detection of aN forces with MFM operated at $\SI{100}{mK}$. Using MFM in frequency-modulated detection mode, force gradient sensitivities down to $\SI{0.14}{\micro N/m}$ have been reported~\cite{straver_phdthesis_2004}. This is well below the expected stray field force gradients $\partial F_y/\partial y=\SI{7.8}{\micro N/m}$ ($d=\SI{20}{nm}$) and $\partial F_y/\partial y=\SI{45}{\micro N/m}$ ($d=\SI{5}{nm}$).

In summary, we have discussed a direct method to detect spin accumulation in a non-magnetic metal strip. The proposed approach is based on the measurement of the magnetic stray field arising from the electron spin accumulation close to the surfaces of the metal strip . To this end, we have derived an analytical expression for the spin accumulation and the corresponding magnetic stray field around a non-magnetic, metallic strip with rectangular cross-section. Based on this, we proposed a sample geometry for a future experiment and calculated the spatial distribution of the magnetic stray field. We showed that the stray field is large enough for detection using the state-of-the-art sensing techniques. Besides, we compared the stray field to the Oersted field around the non-magnetic conductor and found that for the proposed trilayer sample geometry, the Oersted field is dominated by the stray field near the edges of the conducting strip. Such a direct detection of spin accumulation should enable a reliable measurement of spin transport properties, such as spin diffusion length and spin Hall angle, in metals thereby circumventing interfacial complexities.

\section*{Acknowlegdments}
We acknowledge funding from DFG via Priority program 1538 Spin-Caloric Transport (Project GO 944/4) and SPP1601 (HU 1896/2). AK is funded by A. v. Humboldt foundation.

\appendix
\section{Relation between spin chemical potential and spin density}\label{app:spinaccdens}
In order to derive the relation between spin chemical potential and spin density in a non-magnetic conductor, we consider a degenerate non-magnetic gas and obtain relation between spin accumulation and spin density using a two spin channel model. The discussion herein borrows heavily from Ref.\,\citenum{fabian_semiconductor_2007}. For a Fermion gas, we have:
\begin{equation*}
 n = \int g(E) f(E - \mu) dE,
\end{equation*}
where $n$ denotes the density of electrons, $g(E)$ is the density of states per unit volume, $\mu$ is the chemical potential, and $f(x) = 1/(\exp{(x/k_b T)} + 1)$ is the Fermi function. For a two spin model, the above equation becomes:
\begin{equation*}
 n_{\uparrow,\downarrow} = \int g_{\uparrow,\downarrow}(E) f(E - \mu_{\uparrow,\downarrow}) dE.
\end{equation*}
Here, the subscript arrows $\uparrow, \downarrow$ denote two opposite spin direction (up and down, resp.~left and right).
For a non-magnetic conductor, $g_{\uparrow}(E) = g_{\downarrow}(E) = g(E)/2$ with $g(E)$ as the total density of states per unit volume. In addition, we define the following notation:
\begin{align}
 n & =  n_{\uparrow} + n_{\downarrow}, \\
 s & =  n_{\uparrow} - n_{\downarrow}, \\
 \mu & =  \frac{\mu_{\uparrow} + \mu_{\downarrow}}{2}, \\
 \mu_{s} & =  \frac{\mu_{\uparrow} - \mu_{\downarrow}}{2}.
\end{align}
Further we assume, $\mu_s \ll \mu$ for a linear response theory. Having defined the above notation we proceed to express $s$ in terms of spin chemical potential $\mu_s$.
\begin{align}
 s & =  n_{\uparrow} - n_{\downarrow}, \nonumber \\
   & =  \frac{1}{2} \int g(E) ( f(E - \mu_{\uparrow}) - f(E - \mu_{\downarrow}) dE. \label{eq:inter1} \\
\end{align}
We notice the following relations:
\begin{align*}
 \mu_{\uparrow,\downarrow} & =  \mu \pm \mu_s, \\
 \therefore f(E - \mu_{\uparrow,\downarrow}) & =  f(E - (\mu \pm \mu_s)), \\
   & =  f(E - \mu) \pm \frac{\partial f}{\partial \mu} \mu_s, \\
   \therefore f(E - \mu_{\uparrow}) - f(E - \mu_{\downarrow}) & =  2 \frac{\partial f}{\partial \mu} \mu_s.
\end{align*}
Since $\partial f/ \partial \mu$ for a degenerate gas is approximately $\delta(E - \mu)$ \cite{fabian_semiconductor_2007}, we obtain on substitution of the above equation in Eq.~(\ref{eq:inter1}).
\begin{align}
 s & =  \int g(E) \delta(E - \mu) \mu_s dE, \nonumber \\
  & =  g(\mu) \mu_s. \label{eq:smus}
\end{align}
Hence we have a relation between spin density and spin chemical potential via density of states at the chemical potential. It might however be desirable to express the the above relation in terms of commonly used parameters such as conductivity ($\sigma$) and diffusion constant ($D$). This is achieved by comparing the diffusion current formulations used in Refs.~\citenum{chen_theory_2013} and \citenum{fabian_semiconductor_2007}.

The one dimensional particle diffusion current density in the formulation used in Ref.~\citenum{fabian_semiconductor_2007} is given by:
\begin{equation*}
 J_{n} = - D \frac{\partial n}{\partial x},
\end{equation*}
where $D$ is the diffusion constant of the material, and subscript $n$ reminds us that we are talking about a particle current. Correspondingly we can write net spin ``particle'' current:
\begin{align}
 J_{n_\uparrow} - J_{n_\downarrow} & =  - D  \frac{\partial (n_\uparrow - n_\downarrow)}{\partial x},  \nonumber \\
     & =  - D g(\mu) \frac{\partial \mu_s}{\partial x}, \label{eq:jnmus}
\end{align}
where we used Eq.~(\ref{eq:smus}) in the last step above. Using formulation used in Ref.~\onlinecite{chen_theory_2013}, we have
\begin{equation*}
 J_{s}  =  - \frac{\sigma}{2 e} \frac{\partial \mu_s}{\partial x}.
\end{equation*}
Please note that the current density above has been expressed in units of charge current density for convenience \cite{chen_theory_2013}. In order to compare the above expression to Eq.~(\ref{eq:jnmus}), we need to divide the above equation by elementary charge ($e$) throughout. On comparison of the two particle current, we obtain the following relation:
\begin{equation*}
 g(\mu) = \frac{\sigma}{2 e^2 D}.
\end{equation*}
Hence using the above equation in conjunction with Eq.~(\ref{eq:smus}), we obtain the desired relation between spin imbalance density and spin accumulation:
\begin{equation*}
 s = \frac{\sigma}{2 e^2 D} \mu_s.
\end{equation*}

Thus, for the $y$($z$)-polarized electrons, we get
\begin{equation*}
 n_{sy} = \frac{\sigma_N}{2 e^2 D} \mu_{sy}  \qquad\text{and}\qquad  n_{sz} = \frac{\sigma_N}{2 e^2 D} \mu_{sz}
\end{equation*}

\section{Magnetic field of spin accumulation}\label{app:fluxdensity}

The magnetic flux density (in the following referred to as \textit{magnetic field}) originating from a magnetic moment $\mathbf{m}$ is given by\cite{jackson_electrodynamics_1998}
\begin{equation*}
 \mathbf{B} = \frac{\mu_0}{4 \pi r'^3} \left[3 (\mathbf{m} \cdot \hat{\mathbf{r}}') \hat{\mathbf{r}}' - \mathbf{m} \right],
\end{equation*}
where $\mathbf{r}'$ is the position vector from the magnetic moment to the point at which the flux density is calculated.

To obtain the magnetic stray field arising from the spin accumulation in the conducting strip, we integrate the contribution of the magnetic moments within the volume of the strip. With $n_s(\mathbf{r})$ being the spin accumulation density 
at point $\mathbf{r}=(x,y,z)$ with spin polarization $\hat{\mathbf{n}}$, the orientation-dependent magnetic moment density is $\gamma\hbar/2~ \hat{\mathbf{n}} n_s(\mathbf{r})$, where $\gamma$ denotes the gyromagnetic ratio of the material, respectively. Note that we treat the magnetic fields generated by the magnetic moments oriented along $\mathbf{\hat{z}}$ and $\mathbf{\hat{y}}$ initially independently and then calculate the vector sum of the magnetic fields.

We obtain for the magnetic field at $\mathbf{r}_p=(x_p,y_p,z_p)$, caused by the magnetization $\gamma\hbar / 2 \hat{\mathbf{n}} n_s(\mathbf{r})$ present in an infinitesimal volume element $dx\,dy\,dz$ around $\mathbf{r}$
\begin{widetext}
\begin{equation}\label{eq:dB}
d\mathbf{B} = \frac{\mu_0}{8 \pi \left|\mathbf{r}_p-\mathbf{r}\right|^3} \left[3 \frac{\gamma\hbar \hat{\mathbf{n}} n_s(\mathbf{r}) \cdot \left(\mathbf{r}_p-\mathbf{r}\right)}{ \left|\mathbf{r}_p-\mathbf{r}\right|^2} \left(\mathbf{r}_p-\mathbf{r}\right) - \gamma\hbar \hat{\mathbf{n}} n_s(\mathbf{r}) \right]dx\,dy\,dz.
\end{equation}
Employing Eqs.~(\ref{eq:musy}) and (\ref{eq:dB}), we calculate the magnetic field distribution outside the conductor originating from the spin accumulation. Due to the translational symmetry of the problem with respect to the $\mathbf{\hat{x}}$-axis, the magnetic field does not depend on $x_{\text{p}}$ which we choose to be 0.

We begin with the integration along the $\mathbf{\hat{x}}$ and $\mathbf{\hat{y}}$-direction considering only the accumulation of $\mathbf{\hat{y}}$-polarized spins, i.\,e.~the contribution from $n_{sy}$.  This magnetic field contribution is called $\mathbf{B}^{\hat{\mathbf{n}}\parallel \hat{\mathbf{y}}}$ in the following. Note that the integration can be done easily as $n_{sy}$ does not depend on $x$ and $y$. We obtain
\begin{equation}\label{eq:Bypol-xyint}
\int_{x=-\infty}^{\infty} \int_{y=-y_0/2}^{y_0/2} d\mathbf{B}^{\hat{\mathbf{n}}\parallel \hat{\mathbf{y}}}(\mathbf{r},\mathbf{r}_p)
= \frac{\mu_0\gamma\hbar n_{sy}(z)}{8\pi}  
     \left[
     \left( \begin{array}{c}
     0 \\
     \frac{2\left(y_p-\frac{y_0}{2}\right)}{\left(y_p-\frac{y_0}{2}\right)^2+\left(z_p-z\right)^2} \\
     \frac{2\left(z_p-z\right)}{\left(y_p-\frac{y_0}{2}\right)^2+\left(z_p-z\right)^2}
        \end{array} \right) -
     \left( \begin{array}{c}
     0 \\
     \frac{2\left(y_p+\frac{y_0}{2}\right)}{\left(y_p+\frac{y_0}{2}\right)^2+\left(z_p-z\right)^2} \\
     \frac{2\left(z_p-z\right)}{\left(y_p+\frac{y_0}{2}\right)^2+\left(z_p-z\right)^2}
        \end{array} \right)
        \right] dz.
\end{equation}
An analogous integration along the $\mathbf{\hat{x}}$ and $\mathbf{\hat{z}}$-direction for the magnetic field contribution from $n_{sz}$ yields
\begin{equation}\label{eq:Bypol-xzint}
\int_{x=-\infty}^{\infty} \int_{z=-z_0/2}^{z_0/2} d\mathbf{B}^{\hat{\mathbf{n}}\parallel \hat{\mathbf{z}}}(\mathbf{r},\mathbf{r}_p)
 = \frac{\mu_0\gamma\hbar n_{sz}(y)}{8 \pi} 
    \left[
     \left( \begin{array}{c}
     0 \\
     \frac{2\left(y_p-y\right)}{\left(y_p-y\right)^2+\left(z_p-\frac{z_0}{2}\right)^2} \\
     \frac{2\left(z_p-\frac{z_0}{2}\right)}{\left(y_p-y\right)^2+\left(z_p-\frac{z_0}{2}\right)^2}
        \end{array} \right) -
     \left( \begin{array}{c}
     0 \\
     \frac{2\left(y_p-y\right)}{\left(y_p-y\right)^2+\left(z_p+\frac{z_0}{2}\right)^2} \\
     \frac{2\left(z_p+\frac{z_0}{2}\right)}{\left(y_p-y\right)^2+\left(z_p+\frac{z_0}{2}\right)^2}
        \end{array} \right)
        \right] dy.
\end{equation}
\end{widetext}
%
For a full quantitative modelling of the magnetic field distribution in the surrounding of the conductor, we perform the remaining integration over the $y$- and $z$-dimensions of Eqs.~(\ref{eq:Bypol-xyint}) and (\ref{eq:Bypol-xzint}) numerically. To this end, we use the spatially dependent spin accumulation density from Eqs.~(\ref{eq:musy}) -- (\ref{eq:spinaccdensity}).

Before discussing the numerical results below, we turn to a simplified picture where we consider all spins to be located at the conductor's surface (as indicated in Fig.~\ref{fig:Schematic})---a situation with can be treated analytically. This approximation agrees well with the exact solution when the point of interest is located much further away from the conductor as compared to the spin relaxation length ($\sim$ 1 nm for platinum). In this case, we approximate the spin accumulation density for the $\mathbf{\hat{y}}$-polarized electrons as
\begin{equation}
n_{sy}(z) \approx \tilde{n}_{sy} \left[\delta\left(z-\frac{z_0}{2}\right) - \delta\left(z+\frac{z_0}{2}\right)\right]
\end{equation}
with $\delta(x)$ the Dirac delta distribution and $\tilde{n}_{sy}:=\int_0^{z_0/2}n_{sy}(z)\,dz$.
Combining Eqs.~(\ref{eq:musy}) and (\ref{eq:spinaccdensity}), we get
\begin{equation}
\tilde{n}_{sy} = -\frac{j_e \theta\lambda^2}{D e}\left(1-\frac{1}{\cosh\left(\frac{z_0}{2\lambda}\right)}\right)
\end{equation}
Performing the integration for the $z$-direction in Eq.~(\ref{eq:Bypol-xyint}) 
we obtain for the magnetic stray field caused by the spin accumulation $\tilde{n}_{sy}$
\begin{equation*}
\mathbf{B}^{\hat{\mathbf{n}}\parallel \hat{\mathbf{y}}}(\mathbf{r}_p)= \frac{\mu_0 \gamma\hbar \tilde{n}_{sy}}{8 \pi}\mathbf{F}(y_0,z_0;\mathbf{r}_p)
\end{equation*}
with
\begin{widetext}
\begin{multline}
\mathbf{F}(y_0,z_0;\mathbf{r}_p) =
     \left[
     \left( \begin{array}{c}
     0 \\
     \frac{2(y_p-y_0/2)}{(y_p-y_0/2)^2+(z_p-z_0/2)^2} \\
     \frac{2(z_p-z_0/2)}{(y_p-y_0/2)^2+(z_p-z_0/2)^2}
        \end{array} \right) -
     \left( \begin{array}{c}
     0 \\
     \frac{2(y_p+y_0/2)}{(y_p+y_0/2)^2+(z_p-z_0/2)^2} \\
     \frac{2(z_p-z_0/2)}{(y_p+y_0/2)^2+(z_p-z_0/2)^2}
        \end{array} \right)\right.\\
   \left. -
     \left( \begin{array}{c}
     0 \\
     \frac{2(y_p-y_0/2)}{(y_p-y_0/2)^2+(z_p+z_0/2)^2} \\
     \frac{2(z_p+z_0/2)}{(y_p-y_0/2)^2+(z_p+z_0/2)^2}
        \end{array} \right) +
     \left( \begin{array}{c}
     0 \\
     \frac{2(y_p+y_0/2)}{(y_p+y_0/2)^2+(z_p+z_0/2)^2} \\
     \frac{2(z_p+z_0/2)}{(y_p+y_0/2)^2+(z_p+z_0/2)^2}
        \end{array} \right)
        \right]
\end{multline}
\end{widetext}
For the magnetic field contribution of the $z$-polarized electrons, we find correspondingly
\begin{equation*}
\mathbf{B}^{\hat{\mathbf{n}}\parallel \hat{\mathbf{z}}}(\mathbf{r}_p)=-\frac{\mu_0 \gamma\hbar \tilde{n}_{sz}}{8 \pi }\mathbf{F}(y_0,z_0;\mathbf{r}_p)
\end{equation*}
with
\begin{equation}\label{eq:nsytilde}
\tilde{n}_{sz} = \frac{j_e \theta\lambda^2}{D e}\left(1-\frac{1}{\cosh\left(\frac{y_0}{2\lambda}\right)}\right).
\end{equation}

In total, the magnetic field at point $\mathbf{r}_p$ arising from the spin polarization in the conducting strip is given by
\begin{multline}
\mathbf{B}(\mathbf{r}_p)
= \mathbf{B}^{\hat{\mathbf{n}}\parallel \hat{\mathbf{y}}}(\mathbf{r}_p) + \mathbf{B}^{\hat{\mathbf{n}}\parallel \hat{\mathbf{z}}}(\mathbf{r}_p) \\
= \frac{\mu_0\gamma\hbar}{8 \pi} \frac{j_e \theta\tausf}{e} \left(\frac{1}{\cosh\left(\frac{y_0}{2\lambda}\right)} - \frac{1}{\cosh\left(\frac{z_0}{2\lambda}\right)}\right)\mathbf{F}(y_0,z_0;\mathbf{r}_p) \label{eq:B8}
\end{multline}

Obviously, the magnetic stray field is proportional to the spin Hall angle, the spin-flip time and the applied current density through the conductor. Regarding the geometry, a square cross-section of the conductor (i.\,e.~$y_0=z_0$) would imply a vanishing stray field as Eq.~\ref{eq:B8} shows. This is a consequence of the symmetry of the problem and holds for the analytical approximation as well as for the full numerical calculation.
As we are interested in maximizing the stray field around the conductor, we suggest a very thin ($y_0\lesssim 3\,\text{nm}$) metal strip with $z_0\gg y_0$ for the experimental investigation of the calculated stray field. 

\section{Oersted field}\label{app:oerstedfield}
The magnetic field induced by an infinitesimal conductor cross-section $dy\,dz$ around $\mathbf{r}$ can be written as \footnote{we assume the conductor to be aligned along the $x$-axis.}
\begin{equation}
d\mathbf{B}_{\text{oer}}=\frac{\mu_0}{2\pi\left|\mathbf{r}_p-\mathbf{r}\right|^2} \mathbf{j}\times\left(\mathbf{r}_p-\mathbf{r}\right)dy\,dz.
\end{equation}

The total Oersted field arising from the (uniform) current density $\mathbf{j}=\je\hat{\mathbf{x}}$ in the conducting strip can thus be calculated by integrating $d\mathbf{B}_{\text{oer}}$ over the cross-section of the strip. The integral can be solved analytically but the resulting expression is unwieldy and therefore not given here. Figure~\ref{fig:OerstedFieldSLSpatial} shows the spatial distribution of the Oersted field around the conductor. It has its maximum of about $\SI{16}{\micro T}$ at the left and right edge of the strip. For $\rpabs\gg y_0,z_0$, the Oersted field decays proportional to $1/\rpabs$ as expected for the farfield of a current in  a wire. Thus, in the farfield, the Oersted fields dominates the stray field. This is also illustrated in Fig.~\ref{fig:SinglelayerStrayfieldColorPlotAndFarfieldLimits}, where the ratio $|\mathbf{B}|/|\mathbf{B_{\text{oer}}}|$ is plotted as a function of the sensor position $\rp$. Only for small distance from the conducting strip, the stray field exceeds the Oersted field.

Nevertheless, the spatial dependence of the Oersted field significantly differs from that of the magnetic stray field of spin accumulation. Thus, using a spatially resolved magnetic field sensing technique would in principle allow to differentiate between stray field and Oersted field.


%

\end{document}